\documentstyle[12pt]{article}

\begin{document}
\newcommand{\be}{\begin{equation}}
\newcommand{\ben}{\begin{subequations}}
\newcommand{\een}{\end{subequations}}
\newcommand{\beq}{\begin{eqnarray}}
\newcommand{\eeq}{\end{eqnarray}}
\newcommand{\ee}{\end{equation}}
\newcommand{\s}{\\ \vspace*{-4mm}}
\newcommand{\wt}{\widetilde}
\newcommand{\mchi}{\mbox{$m_{\tilde {\chi}_1^0}$}}
\newcommand{\epem}{\mbox{$e^+ e^-$}}
\newcommand{\mchisq}{m_{\tilde {\chi}_1^0}^2}
\newcommand{\lsp}{\mbox{$\tilde {\chi}_1^0$}}
\newcommand{\tanb}{\mbox{$\tan \! \beta$}}
\renewcommand{\thefootnote}{\fnsymbol{footnote}}

\begin{flushright}
TUM--HEP--398--00 \\
LC--TH--2001--013 \\
January 2001\\
\end{flushright}

\vspace*{2.5cm}

\begin{center}
{\Large \bf  Dark Matter and the SUSY Mass Scale\footnote{To appear
in the Proceedings of the {\it ECFA--DESY Workshop on Physics Studies
for a Future Linear Collider}.}} \\
\vspace{5mm}
Manuel Drees \\
\vspace{2.5mm}
{\it Physik Department, TU M\"unchen, D--85748 Garching, Germany}

\end{center}
\vspace{10mm}

\begin{abstract}
The connection between the present density of neutralinos that are
left over from the Big Bang and the superparticle mass scale is
briefly reviewed. Superparticle mass scales in the range from a few
GeV to several TeV can lead to an acceptable density of thermal relic
neutralinos, the actual value depending on relations between the
masses of certain sparticles and Higgs bosons.

\end{abstract}
\clearpage

\bigskip
\noindent
Most theoretical physicists believe that some ``new physics'' will
have to appear at or below the TeV scale, in order to render the
scalar sector of the SM (technically) natural. This consensus
motivates the collider physics community to design and (hopefully)
build particle colliders that can directly probe TeV
energies. However, while from a string theorist's perspective there
would be little difference between ``new physics'' scales of 300 GeV
and 1 TeV, it is clear that from the collider physicist's point of view
a more precise statement regarding the energy needed to unravel the
mystery shrouding the scalar sector of the SM would be invaluable.

This issue is most frequently discussed in the framework of
supersymmetric theories, which have been worked out in much more
detail than any competing theories. There have been various attempts
to make the naturalness argument more precise by defining quantitative
measures of finetuning, at least in the (very attractive) class of
models where the electroweak gauge symmetry is broken radiatively
\cite{dm}. However, even if one of these definitions is accepted, one
still has to use one's judgment as to how much finetuning one is
willing to tolerate.

Calculations of the density of thermal Big Bang relics seem to allow
to derive more precise bounds on sparticle masses, if we require that
relic LSPs (usually assumed to be the lightest neutralino \lsp) have
just about the right density indicated by cosmological observations
\cite{primack}. It should be clear from the start that such arguments
do not really touch the main motivation for postulating the existence
of superparticles at the weak scale. After all, if we learned tomorrow
that the Dark Matter in the Universe consists of axions, few people
would conclude that weak--scale supersymmetry has been ruled out!

Let us nevertheless press on and explore the consequences of requiring
\lsp\ to form the Dark Matter. One immediate requirement is that it
must be stable. This excludes models with broken $R$ parity (where
\lsp\ decays into SM particles) as well as models with gauge mediated
SUSY breaking (where \lsp\ decays into a gravitino and a photon, if it
is the lightest visible sector superparticle).\footnote{In other
words, any SUSY model can be made ``Dark Matter safe'' by making \lsp\
unstable. This can be accomplished without modifying any collider
signatures, e.g. by introducing a tiny amount of $R-$parity breaking,
or by letting \lsp\ decay into a hidden--sector superparticle. Of
course, one then has to find another explanation for the Dark Matter.}
In order to make quantitative statements, we have to assume in
addition that the post--inflationary Universe was hot enough for \lsp\
to have been in chemical equilibrium with SM particles, i.e. that the
rate for reactions that create or destroy superparticles was higher
than the expansion rate. This typically requires the
post--inflationary reheat temperature $T_R$ to exceed $\sim 10\%$ of
\mchi. Note that we currently only know that $T_R \geq 1$ MeV, since
otherwise nucleosynthesis could not have occurred \cite{kolb}. Given
that the inflaton mass is supposed to be around $10^{13}$ GeV,
assuming $T_R \geq \mchi/10$ is not unreasonable, but models can be
constructed where this condition is not satisfied.\footnote{Indeed,
most SUSY models need $T_R \ll 10^{13}$ GeV in order to avoid
over--production of gravitinos.}

Given the assumption of chemical equilibrium, the present \lsp\ relic
density can be computed quite reliably, if the sparticle and Higgs
spectrum is known. Not surprisingly, one finds that the relic density
is essentially proportional to the inverse of the cross section for
\lsp\ annihilation into SM particles. In general \lsp\ is a linear
superposition of the bino (the superpartner of the $U(1)_Y$ gauge
boson), the neutral wino (the superpartner of the neutral $SU(2)$
gauge bosons), and the two neutral higgsinos. These interaction
eigenstates receive a priori unknown masses $M_1, \ M_2$ and $\mu$,
respectively, while mixing between these states is induced by
off--diagonal mass terms ${\cal O}(M_Z)$. In most models with
(approximate) gaugino mass unification and radiative gauge symmetry
breaking, the LSP is bino--like. This follows from the large size of
the top Yukawa coupling, which drives the squared mass of the Higgs
boson that couples to top quarks to too negative a value, unless it
receives a large positive contribution $\mu^2$. Note also that RG
running from the GUT to the weak scale reduces the bino mass by about
a factor of 2.5. These effects together imply that usually $|M_1| <
|M_2| \leq |\mu|$, leading to a bino--like LSP, independent of details
of the scalar spectrum at the GUT scale \cite{toby}. In this case the
LSP relic density can usually be estimated from $\lsp \lsp \rightarrow
\ell^+ \ell^-$ annihilation ($\ell=e,\mu,\tau$) through the exchange
of $SU(2)$ singlet sleptons $\tilde{\ell}_R$ in the $t-$ or
$u-$channel. The reason is that $\tilde{\ell}_R$ has the largest
hypercharge of all sfermions; in most models it is also among the
lightest of all sfermions. The scaled LSP relic density multiplied
with the scaled Hubble constant can then be estimated as \cite{dn3}
\be \label{e1}
\Omega_{\tilde \chi_1^0} h^2 
\simeq \frac
{\left( m^2_{\tilde{\chi}_1^0} + m^2_{\tilde{\ell}_R} \right)^4} 
{10^6 \ {\rm GeV}^2 m^2_{\tilde{\chi}_1^0} \left(
m^4_{\tilde{\ell}_R} + m^4_{\tilde{\chi}_1^0} \right) }
\ee
{\em If} eq.(\ref{e1}) is valid, it is easy to see that the
cosmological constraint \cite{primack} $\Omega_{\tilde{\chi}_1^0} h^2
\leq 0.3$ requires $\mchi, m_{\tilde{\ell}_R} \leq 200$ GeV. This
argument is independent of details of the Higgs sector (unless $2
\mchi \simeq m_{\rm Higgs}$; see below). It is encouraging that
eq.(\ref{e1}) leads to sparticle masses in the few hundred GeV range,
where one would expect them from naturalness arguments.\footnote{A
much lighter sparticle spectrum can also lead to a cosmologically
interesting relic density, $\Omega_{\tilde{\chi}_1^0} h^2 \sim 0.1$,
if $\mchi \ll m_{\tilde{\ell}_R}$. Of course, such light spectra are
nowadays excluded by LEP searches.} This is quite nontrivial, since
the numerical constant appearing in this equation depends on
quantities like the Planck mass and the temperature of the cosmic
microwave background.

However, in general things are not so simple, which is why I used
qualifiers like ``in most models'' and ``usually'' in the previous
paragraph. Perfectly acceptable SUSY models exist where \lsp\ is {\em
not} bino--like; in other cases, \lsp\ is bino--like, but
eq.(\ref{e1}) overestimates the true thermal relic density by several
orders of magnitude. The ``bounds'' $\mchi, m_{\tilde{\ell}_R} \leq
200$ GeV therefore have several loopholes:
\begin{itemize}
\item Even in mSUGRA, i.e. if strict universality is imposed at the
GUT scale $M_X = 2 \cdot 10^{16}$ GeV, \lsp\ can have a large or even
dominant Higgsino component \cite{feng}, if the scalar mass $m_0 \gg$
the gaugino mass $M_{1/2}$. Since higgsinos annihilate quite
efficiently into $W$ and $Z$ pairs, the upper bound on \mchi\ then has
to be raised to $\sim 1.5$ TeV \cite{gondolo}. Even worse, {\em no}
upper bounds on gaugino or sfermion masses can be given in this case,
as long as $m_0 \gg M_{1/2}$ holds.\footnote{I should emphasize that
there are also solutions with $m_0 \gg M_{1/2}$ where \mchi\ is in the
(few) hundred GeV range and the relic density is acceptable, if both
the higgsino and the gaugino components of \lsp\ are sizable. In fact,
this region of parameter space is quite favorable from the point of
view of Dark Matter searches \cite{feng2}. Here the light chargino and
lighter neutralinos should be accessible at future \epem\ linear
colliders, but sleptons would be out of reach. See ref.\cite{feng_new}
for further discussion of these solutions.} Gaugino mass unification
implies that the gluino mass $m_{\tilde g} \geq 6 \mchi$. This becomes
a strong inequality in the Higgsino region, where $|M_1| > |\mu|$. A
1.5 TeV higgsino--like LSP thus implies a gluino mass of at least 10
TeV. Since $m_0 \gg M_{1/2}$, the squarks would be even heavier.  If
the soft breaking Higgs masses exceed squark masses at the GUT scale
\cite{dkn}, or if the SUSY messenger scale is significantly below the
GUT scale \cite{spaniards}, the higgsino component of \lsp\ increases,
i.e. a higgsino--like LSP becomes possible for smaller ratios of
sfermion and gaugino masses (but the ratio of gluino and LSP masses
remains unchanged).
\item If the $SU(2)$ gaugino mass $M_2 <$ the $U(1)_Y$ gaugino mass
$M_1$ at the weak scale, \lsp\ will be wino--like rather than
bino--like. Winos annihilate even more efficiently into $W$ and $Z$
pairs than higgsinos do (since they are $SU(2)$ triplets, rather than
doublets), so that LSP masses up to $\sim 2$ TeV become
acceptable. Again, there is {\em no} bound on sfermion masses in this
case. Such a scenario is e.g. realized in anomaly--mediated SUSY
breaking. Anomaly mediation predicts $m_{\tilde g} \simeq 10 \mchi$,
so that gluino masses well above 10 TeV are again cosmologically
acceptable. 
\item Even if \lsp\ is bino--like, eq.(\ref{e1}) might be wildly off
the mark. \lsp\ pair annihilation is enhanced by several orders of
magnitude if $2 \mchi \simeq m_A$, where $m_A$ is the mass of the
CP--odd Higgs boson of the MSSM \cite{dn3}. In models with radiative
gauge symmetry breaking, including mSUGRA, this can happen if the
ratio of vevs \tanb\ is large. In this case $\mchi \simeq 1$ TeV can
be allowed, independent of the values of the sfermion masses. Even
heavier LSPs can be cosmologically acceptable if $\mchi \simeq
m_{\tilde{t}_1}$, since then $\lsp - \tilde{t}_1$ co--annihilation
reduces the relic density by up to three orders of magnitude
\cite{celine}; here $\tilde{t}_1$ is the lighter scalar top mass
eigenstate. Finally, if $\mchi \simeq m_{\tilde{\tau}_1}, \
\Omega_{\tilde{\chi}_1^0} h^2$ is reduced by up to one order of
magnitude \cite{ellis}, i.e. the upper bounds on \mchi\ and
$m_{\tilde{\ell}_R}$ have to be increased by about a factor of 3. In
this case a 1.5 TeV lepton collider, and the LHC, would still be
guaranteed to see a SUSY signal \cite{ellis2}. Within mSUGRA this
particular loophole might be the most ``likely'' one, since one
doesn't need large ratios of soft breaking parameters (as in the $\lsp
\simeq \tilde{h}$ loophole), nor does one need $\tanb \gg 1$ (as in
the $2 \mchi \simeq m_A$ loophole). However, the ratio $m_0 / M_{1/2}$
has to be within $\sim 5\%$ of its lower bound, which is set by the
requirement $\mchi < m_{\tilde{\tau}_1}$.

\end{itemize}

In summary, neutralino Dark Matter seems to be most natural if $\mchi,
m_{\tilde{\ell}_R} \leq 200$ GeV. However, several loopholes exist
that allow $\mchi \geq 1$ TeV ($\lsp \simeq \tilde{h}, \lsp \simeq
\wt{W}_3, 2 \mchi \simeq m_A, \mchi \simeq m_{\tilde{t}_1}$), while
$\mchi \simeq m_{\tilde{\tau}_1}$ allows \mchi\ up to $\sim 600$
GeV. Assessing the probability that Nature chose one of these
loopholes is very difficult and model--dependent. Deviating from the
``canonical'' mSUGRA framework can make things worse (e.g., $\lsp
\simeq \wt{W}_3$ is almost automatic in models with anomaly mediated
SUSY breaking) or better (e.g. in SUSY GUTs \cite{hitoshi} or models
with intermediate $SU(4) \times SU(2)_L \times SU(2)_R \times U(1)$
symmetry, where $\mchi \simeq m_{\tilde{\tau}_1}$ becomes more
difficult to realize). Of course, the fact that cosmologically
acceptable models can be constructed where neither the LHC nor even a
5 TeV lepton collider would detect a SUSY signal doesn't mean that
such models are ``natural''.

My personal conclusion is that {\em if} the lightest neutralino is
stable, {\em and if} it was in chemical equilibrium in the
post--inflationary Universe, the requirement
$\Omega_{\tilde{\chi}_1^0} h^2 < 0.3$ excludes large regions of SUSY
parameter space with $\mchi > 200$ GeV or $m_{\tilde{\ell}_R} > 200$
GeV. However, given the assumptions needed to derive these ``bounds'',
and the numerous loopholes that permit much heavier LSPs without
``overclosing the Universe'', this cosmological consideration should
probably be viewed as another naturalness argument, independent from
but by no means superior to the arguments based on analyses of
electroweak gauge symmetry breaking.

\bigskip

\noindent {\bf Acknowledgements:}
This work was supported in part by the ``Sonderforschungsbereich
375--95 f\"ur Astro--Teilchenphysik'' der Deutschen
Forschungsgemeinschaft.

\end{document}